\documentclass[12pt]{article}
\usepackage[dvips]{graphicx}
\usepackage{amssymb}
\usepackage{amsmath}
\usepackage{amsfonts}
\usepackage{theorem}
\newtheorem{theor}{Theorem}

\newtheorem{cor}{Corollary}
\newtheorem{rem}{Remark}
\usepackage{hhline}

\begin{document}
\title{The boundary layer problem in Bayesian adaptive quadrature}
\author{Gh.~Adam (1 and 2), S.~Adam (1 and 2)\\
((1) LIT-JINR Dubna Russia,\\
(2) IFIN-HH Magurele--Bucharest Romania)}
\maketitle
\begin{abstract}
The boundary layer of a finite domain $[a, b]$ covers
mesoscopic lateral neighbourhoods, inside $[a, b]$, of the endpoints
$a$ and $b$. The correct diagnostic of the integrand behaviour at
$a$ and $b$, based on its sampling inside the boundary layer, is the first
from a set of hierarchically ordered criteria allowing \emph{a priori}
Bayesian inference on efficient mesh generation in automatic adaptive
quadrature.
\end{abstract}
\section{Introduction}
The boundary layer problem in numerical quadrature is the first, and
probably the most difficult to solve, from a Bayesian chain aiming either
to implement the automatic mesh generation by the adaptive quadrature
methods under sound expectation of \emph{reliable local quadrature rule}
$\{q, e\}$ outputs, or to provide early detection of the origins
of code failure.
\par
The need of such a stringent requirement, which goes well beyond the usual
implementation of the automatic adaptive quadrature rules (see, e.g.,
\cite{QUADPACK,DavRab,KU98})
arises in the numerical exploration of the predictions of models describing
phase transitions in complex physical systems, which critically depend on
the realization of (unknown in advance) values of some specific parameters
(like, e.g., the hole or electron doping level in high-$T\sb{c}$
superconducting materials
\cite{PlSpr}).
The solution of the resulting parametric integrals makes use of the existing
library programs the reliability of the outputs of which is heavily based on
user's ability to choose the suitable procedure from a proposed menu.
The impossibility to know in advance the detailed behaviour of the integrand
function over the whole class of parametric integrals forces the use of a trial and error approach which may result in unnoticed unreliable $\{q, e\}$
pairs and, consequently, bad output failure.
\par
The \emph{a posteriori} assessment of the reliability of the $\{q, e\}$ pair
over the current integration subrange
\cite {AA01,AAP03}
solves only half of the problem since the code remains highly inefficient.
The \emph{a priori} verification of the conditioning of the integrand
profile at the quadrature knots, which was proposed by us some time ago
\cite{AA04}
needed in fact, a whole set of hierarchically ordered criteria providing
\emph{Bayesian inference}
\cite{Jaynes86}
on the status of the \emph{gradually generated\/} integrand profile at
newly added quadrature knots.
\par
The \emph{root} of the resulting Bayesian inference decision tree
is the \emph{diagnostic of the behaviour of the integrand function
$f(x)$ at the boundaries $a$ and $b$ of the finite integration domain
$[a, b]$}.
To set it, a suitable integrand sampling is required inside a
\emph{mesoscopic neighbourhood of the boundary layer of} $[a, b]$.
\par
The present paper generalizes the analysis done in
\cite{AATN05}
over minimal four point partitions inside the mesoscopic regions associated
to each of the ends $a$ and $b$ of $[a, b]$.
By allowing an \emph{unrestricted\/} four point partition, the diagnostic
failures stemming from the inadequacy of a frozen integrand sampling are
avoided, resulting in analysis reliability enhancement for difficult integrand
functions.
\section{Diagnostics and Bayesian inferences from the boundary layer analysis}
Let $x\sb{r}$ denote either $fl(a)$ or $fl(b)$, the floating point
representations of the endpoints $a$ and $b$ of $[a, b]$.
The analysis establishes the status of the value
$f\sb{x\sb{r}} = fl(f(x\sb{r}))$, the \emph{floating point
representation of the computed value of the integrand},
$f(x\sb{r}) \in \mathbb{R}$, as follows:
\begin{itemize}
  \item [(i)] {\emph{Diagnostic: Smooth integrand behaviour}.\\
               \emph{Bayesian inference}: \emph{Regular} quadrature knot
                mesh over $[a, b]$ can be generated starting from the
                $x\sb{r}$ endpoint.\\
               \emph{Supplementary information}: The analysis also generates
                an estimate of the lateral derivative $f'(x\sb{r})$ inside
                $[a, b]$.
                This will serve to the formulation of acceptance check
                criteria for the two quadrature knots which lie nearest and
                next nearest to $x\sb{r}$. In case of rejection, the decision
                to perform the \emph{immediate subrange subdivision of
                $[a, b]$}, without generating the integrand profile at the
                remaining quadrature knots is taken.}
  \item[(ii)] {\emph{Diagnostic:
                Endpoint or outer singularity of $f(x)$ or its derivatives}.\\
               \emph{Bayesian inference}: \emph{Slow convergence} is to be
                expected under subrange subdivision.
                Use of a specific subrange subdivision procedure based on
                bisection together with a convergence
                acceleration procedure (e.g., the epsilon extrapolation
                algorithm) are a must.}
  \item[(iii)] {\emph{Diagnostic:
                Inner nearby singularity}.\\
               \emph{Bayesian inference}: The occurrence of an offending
                inner singular point $x\sb{0}$ near $x\sb{r}$ is to be
                further confirmed.
                Under affirmative diagnostic, $x\sb{0}$ is to be located to
                machine accuracy. The input integral is then split into two
                integrals, over $[a, x\sb{0})$ and $(x\sb{0}, b]$, each
                being further processed following the procedure adequate
                for endpoint singularity.}
  \item[(iv)] {\emph{Diagnostic: Inner nearby finite jump.}\\
               \emph{Bayesian inference}: Further confirmation of the
                existence of an offending inner jump point $x\sb{0}$
                near $x\sb{r}$ is necessary. Under
                affirmative diagnostic, $x\sb{0}$ is to be located to
                machine accuracy. The input integral is then split into two
                integrals, over $[a, x\sb{0}\sp{-})$ and $(x\sb{0}\sp{+}, b]$,
                with $f(x\sb{0}\sp{\mp})$ and $f'(x\sb{0}\sp{\mp})$ taking
                values equal to the lateral limits of $f(x)$ and $f'(x)$ at
                $x\sb{0}$. The resulting integrals over the two subintervals
                are to be solved for smooth $f(x)$.}
  \item[(v)] {\emph{Diagnostic: Irregular behaviour}.\\
              \emph{Bayesian inference}: The output of the automatic procedure
               could hardly be taken for reliable. Clarification of the
               offending integrand behaviour is a must.}
  \item[(vi)] {\emph{Diagnostic: Smooth integrand behaviour at both ends
                $a$ and $b$}.\\
              \emph{Bayesian inference}: Early check for the presence of
               an \emph{oscillatory or odd integrand} is useful.
               If an affirmative diagnostic is issued, then \emph{define
               a ceiling accuracy of the expected output}, originating in
               severe precision loss due to heavy cancellation by subtraction.}
\end{itemize}
\section{Unrestricted least squares analysis}
\begin{theor}
The function $f\! :\! [a, b]\subset\mathbb{R}\rightarrow\mathbb{R}, f=f(x)$,
is \emph{smooth\/} inside a lateral mesoscopic neighbourhood
$V(x\sb{r})\subseteq [a, b]$ of the reference abscissa $x\sb{r}$ denoting
the floating point representation of either the end $a$ or the end $b$ of
$[a, b]$, provided the computed values of the first order divided
differences of $f(x)$ over any abscissa sampling inside $V(x\sb{r})$ are
\emph{independent on the choice of the sampling abscissas}.
\label{theor:smooth}
\end{theor}
\begin{rem}
The result stated in Theorem~\ref{theor:smooth} essentially follows from
the property that, for any reference abscissa $x\sb{r}\in D\subset\mathbb{R}$
of a continuous twice differentiable function
$f\! :\! D\rightarrow\mathbb{R}, f=f(x)$, a nonvanishing neighbourhood
$V(x\sb{r})\subseteq D$ does exist inside which the \emph{linear} Taylor
series expansion of $f(x)$ around $f(x\sb{r})$ holds true within some
predefined accuracy threshold\/ $0<\varepsilon\ll 1$.
\label{rem:taylor}
\end{rem}
The numerical check of the continuity of $f(x)$ at the ends of the
integration domain $[a, b]$ can only be done from a
\emph{ sampling of its computed values},
$\{f\sb{i} = fl(f(x\sb{i})) |i = 0, 1, \cdots, m\}$, over a set of $m+1$
machine number arguments
$S\sb{m}(x\sb{r})\! =\! \{x\sb{i}\!
\in\! V(x\sb{r})  | i\! =\! 0, 1, \cdots, m\}$,
$x\sb{r} \in S\sb{m}(x\sb{r})$, $m\geq 3$.
If $\{f(x\sb{i})) |i = 0, 1, \cdots, m\}$ denote the
\emph{actual\/} values of $f(x\sb{i})$ over $S\sb{m}(x\sb{r})$, then,
due to the round-off, $f(x\sb{i}) - f\sb{i} \neq 0$ in general.
As a consequence, 
the best information on the smoothness properties of
$f(x)$ at $x\sb{r}$ following from the set
$\{x\sb{i}, f\sb{i}\}$ is obtained from the scrutiny of the properties
of a second degree polynomial least squares fit to the floating point data.
\par
A problem in terms of \emph{machine number abscissas} is obtained by
the scale transformation
\begin{equation}
   x\sb{i} = x\sb{0} + \xi\sb{i} h\sb{r}, \quad i = 0,1, \cdots, m; \quad
             \xi\sb{i} \in \mathbb {Z}; \quad \xi\sb{0}=0,
   \label{eq:csii}
\end{equation}
where $h\sb{r}$ denotes the algebraic distance from
$x\sb{r}$ to its nearest machine number inside $[a, b]$.
This leads to the second degree fitting polynomial
\begin{equation}
   y\sb{2}(x\sb{i}) = \alpha\sb{0}\pi\sb{0}(\xi\sb{i}) +
                      \alpha\sb{1}\pi\sb{1}(\xi\sb{i}) +
                      \alpha\sb{2}\pi\sb{2}(\xi\sb{i})\, ,
   \label{eq:y2fit}
\end{equation}
spanned by the \emph{orthonormal\/} basis polynomials
$\pi\sb{k}(\xi\sb{i}), \ k=0,1,2$.
\par
A \emph{linear} polynomial fitting over $S\sb{m}(x\sb{r})$ is obtained
provided $\alpha\sb{2}\pi\sb{2}(\xi)$ is \emph{negligible everywhere}
at $\xi\in [\xi\sb{min}, \xi\sb{max}]$,
$\xi\sb{min}=\min\{\xi\sb{0}, \xi\sb{1}, \cdots, \xi\sb{m}\}$,
$\xi\sb{max}=\max\{\xi\sb{0}, \xi\sb{1}, \cdots, \xi\sb{m}\}$.
Making all calculations requested by the least squares procedure then
results in the statement of the theorem, QED.
\begin{cor}
If we assume a \emph{minimal mesh sampling} $S\sb{3}(x\sb{r})$ characterized
by the abscissa set $\xi\sb{0} = 0, \ \xi\sb{1} = p > 0,\
                     \xi\sb{2} = \mu p > \xi\sb{1}, \
                     \xi\sb{3} = q, \ |\xi\sb{3}| \ll \xi\sb{1}$,
then the following \emph{smoothing criteria} emerge:
  \begin{equation}
     d\sb{20}-\mu d\sb{10}\approx 0, \quad
     \mu d\sb{30}-\frac{q}{p}d\sb{20}\approx 0, \quad
     \frac{q}{p}\Bigr[\frac{q}{p}d\sb{10}-d\sb{30}\Bigl]\approx 0, \quad
     d\sb{ij}=f\sb{i}-f\sb{j}.
  \label{eq:crit1}
  \end{equation}
\label{cor:crita}
\end{cor}
\vspace*{-5mm}
\begin{cor}
If we assume a \emph{minimal mesh sampling} $S\sb{3}(x\sb{r})$ characterized
by the abscissa set $\xi\sb{0} = 0, \ \xi\sb{1} = p\sim\xi\sb{2} = q, \ 
              \xi\sb{3} = r, \ \max\{\xi\sb{1},\xi\sb{2}\} \ll \xi\sb{3}$,
then the following \emph{smoothing criteria} emerge:
  \begin{equation}
    pd\sb{20}-qd\sb{10}\approx 0, \quad
    qd\sb{30}-rd\sb{20}\approx 0, \quad
    rd\sb{10}-pd\sb{30}\approx 0.
  \label{eq:crit2}
  \end{equation}
\label{cor:critb}
\end{cor}
\vspace*{-5mm}
\begin{cor}
 If the analysis issued the diagnostic of smooth $f(x)$ at the endpoint
$x\sb{r}$, then the following estimate for the lateral first order derivative
$f'(x\sb{r})$ inside $[a, b]$ holds,
  \begin{eqnarray}
     &&f'(x\sb{r})\approx\frac{1}{(m+1)\bar{\delta\sp{2}}h\sb{r}}
                       \sum\sb{i=1}\sp{m}\delta\sb{i}d\sb{i0},
  \label{eq:fprim}\\
     &&\delta\sb{i}=\xi\sb{i}-\bar\xi , \quad
     \bar\xi=\frac{1}{m+1}\sum\sb{i=0}\sp{m}\xi\sb{i}, \quad
     \bar{\delta\sp{2}}=\frac{1}{m+1}\sum\sb{i=0}\sp{m}\delta\sb{i}\sp{2}.
  \label{eq:deli}
  \end{eqnarray}
\label{cor:deriv}
\end{cor}
\section{Diagnostic uniqueness}
To allow useful inferences, the analysis done in the previous section
has to be supplemented, on one side, with a quantitative measure of the
smallness of the differences defined in Corollaries~\ref{cor:crita}
and~\ref{cor:critb} and, on the other side, with qualitative criteria
able to single out those specific unique features which characterize
other integrand behaviours inside a mesoscopic neighbourhood of $x\sb{r}$.
\par
{\bf 4.1. Smoothness threshold.}
The practical implementation of the smoothness criteria~(\ref{eq:crit1})
and~(\ref{eq:crit2}) compares the magnitudes of the expressions entering
the left hand sides of these equations with an \emph{integrand dependent
upper threshold},
\begin{equation}
   0 < \tau\sb{f} = \tau\sb{0}\varepsilon\sb{0}
                    \max\{|f\sb{0}|, |f\sb{1}|, |f\sb{2}|, |f\sb{3}|\},
\label{eq:tauf}
\end{equation}
where $\varepsilon\sb{0}$, the machine epsilon with respect to addition,
defines the machine accuracy dependent parameter of the solution, while
$\tau\sb{0}$, $0 < \tau\sb{0}\ll \varepsilon\sb{0}\sp{-1/2}$, is a
heuristic parameter intended to overcome possible diagnostic errors coming
from the normal roundoff noise. In practice, we have found that a value
$\tau\sb{0}=2\sp{10}$ resulted in adequate diagnostics for all the tested
smooth case study functions $f(x)$.
\par
The diagnostic reliability decreases as long as the magnitudes of one or
more of the quantities entering the Eqs.~(\ref{eq:crit1}) or~(\ref{eq:crit2})
get near $\tau\sb{f}$.
\par
{\bf 4.2. Endpoint or outer singularity.}
The basic features of such a behaviour are: inward monotonically decreasing
$|f(x)|$ under $d\sb{i,i-1}\cdot d\sb{i+1,i} > 0$; steep variation of
$f(x)$ over argument distances separated by one or a few machine numbers;
inward monotonically decreasing $|f'(x)|$.
\par
If the sampling of Corollary~\ref{cor:crita} is generated, then the addition
of two supplementary abscissas at $(\xi\sb{1}+\xi\sb{3})/2$ and
$(\xi\sb{1}+\xi\sb{2})/2$ respectively and the confirmation of the
abovementioned three features once again over the resulting smaller subranges
distinguishes singular behaviour from an inner finite jump near $x\sb{r}$.
\par
{\bf 4.3. Inner nearby singularity.}
The features of this kind of behaviour are mirror reflected with respect to
those of an endpoint or outer singularity.
\par
{\bf 4.4. Inner nearby finite jump.}
There are two hints suggesting this diagnostic. First, the sharp increase
of the magnitude of one of the first order divided differences approximating
$f'(x)$ over the subranges defined by sampling. Second, the feature gets
enhanced if a finer partition is defined over the subrange in question.
\par
{\bf 4.5. Irregular behaviour.}
This negative diagnostic is usually associated with the occurrence of rapid
oscillations of $f(x)$ which make the usual local quadrature rules ineffective.
\section{Code robustness: hardware and software environment}
The analysis described above yields diagnostics issued by a code which
runs within an environment defined by the hardware and the software at hand.
The \emph{code robustness, reliability, and portability} are secured provided
several delicate points are adequately solved.
\par
The last code versions were run on several PCs with Intel 4+, AMD32, or
AMD64 processors, Linux 2.4 or Linux 2.6+ operating systems, and the
GNU gcc compiler incorporating Fortran 77. Early code versions were also
run on SUN workstations in BLTP-JINR, or under Microsoft XP OS. 
\par
The study of the conformity of the hardware and software of the abovementioned
systems to the \emph{IEEE 754 standard} which governs the floating
\begin{figure}[ht]
\begin{center}
\resizebox{\textwidth}{!}
{\includegraphics{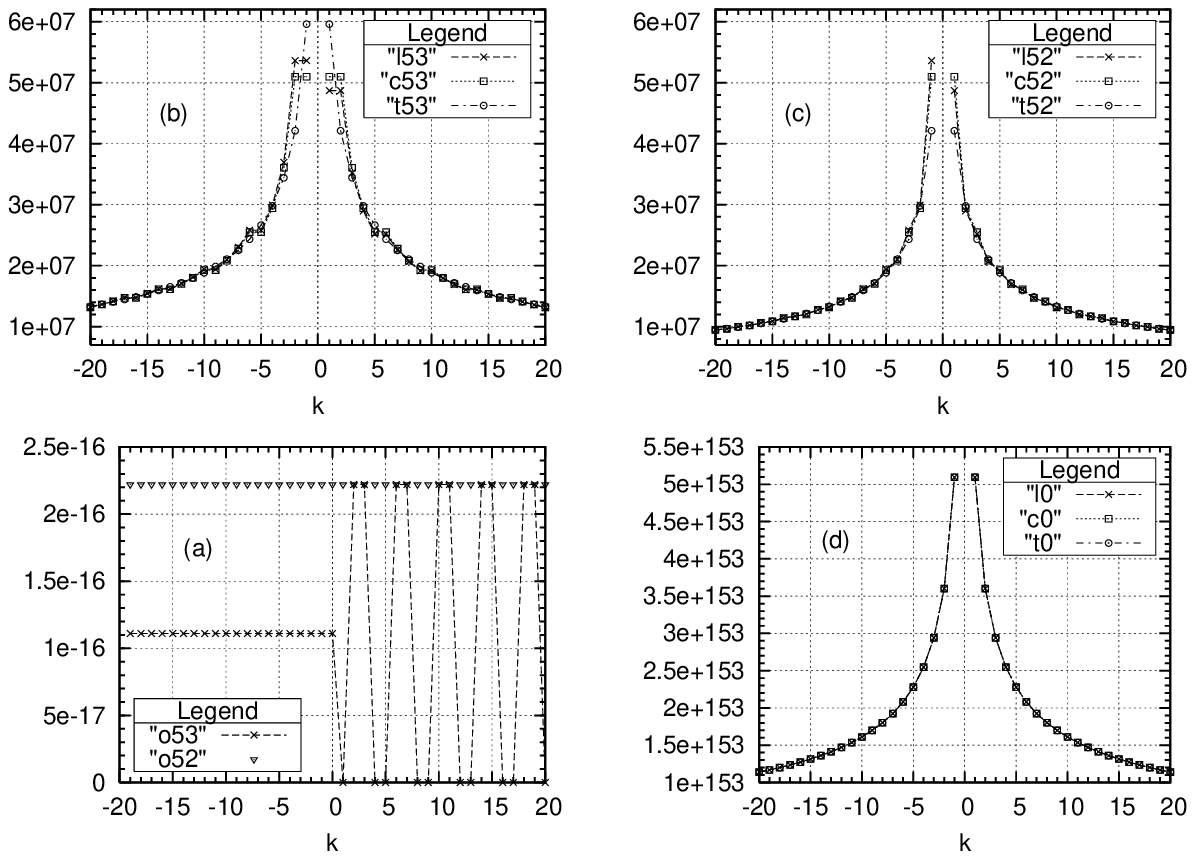}}
\end{center}
\caption{\small Influence of machine epsilon definition on output reliability.
  (a) Check of the fulfilment of the reliability criterion
  $fl(1+(k+1)\varepsilon\sb{0})-fl(1+k\varepsilon\sb{0})=\varepsilon\sb{0}$.
  (Data set "o53" shows that our system is not compliant to the IEEE 754
   standard value $\varepsilon\sb{0}=2\sp{-53}$; data set "o52" shows that
   $\varepsilon\sb{0}=2\sp{-52}$ is the right choice.)
  (b)--(d) Comparison of two code outputs with the true results for the case
   study function~(8):
  (b) -- spuriously equal function pair values occur under
   $\varepsilon\sb{0}=2\sp{-53}$ at arguments intended to represent
   neighbouring machine numbers;
  (c) -- the spuriously equal pairs disappear, but heavy precision loss and
   output precision path dependence are still present near the singular point
   $x\sb{p} = \sqrt{3}-1$ under $\varepsilon\sb{0}=2\sp{-52}$;
  (d) -- all difficulties are solved if singularity is moved to the origin,
   where $\varepsilon\sb{0}=u=2\sp{-1023}$.
  (Legend prefixes: "t" -- \emph{true} results;
   "c" -- \emph{unique} floating point value of $x\sb{p}$ used inside all
   procedures;
   "l" -- calculation of $f(x)$ is done using a \emph{locally
   processor produced approximation} for $x\sb{p}$, which is more accurate
   than that transferred inbetween other procedures.)
}
\label{fig:1}
\end{figure}
point arithmetic (see, e.g.,
\cite{goldb,kah1}),
revealed the occurrence of four instances where the requirements of the
standard were more or less frequently infringed:
(i) \emph{length of the significand;} (ii) \emph{floating point comparisons;}
(iii) \emph{code optimization;} (iv) \emph{underflow threshold}.
\par
{\bf 5.1. Length of the floating point double precision significand.}
Under the standard compliant value of the machine epsilon with respect to
addition, $\varepsilon\sb{0}=2\sp{-53}$, often analysis failure has been
noticed. This was identified to stem from the \emph{loss of the last,
assumed significant, 53-rd bit\/} under reuse of \emph{exact\/}
data after their RAM/cache storage.
\par
The simplest possible case study of hidden bit loss is illustrated in
Fig.~1a: the quantities $a\sb{k}=1+k\varepsilon\sb{0}$,
$\varepsilon\sb{0}=2\sp{-53}$, $k=-20(1)20$, are computed and stored in
an array (to force RAM/cache storage) and then the differences
$\delta\sb{k+1,k}=a\sb{k+1}-a\sb{k}$ are plotted. Under IEEE 754 standard
observance, the \emph{constant answer},
$\ \delta\sb{k+1,k}=\varepsilon\sb{0}\ $ should had been obtained.
However, this result is obtained indeed under $\varepsilon\sb{0}=2\sp{-52}$.
\par
Fig.~1b illustrates the same effect when exploring integrand behaviour
around a singular point on an example taken from QUADPACK,
\cite{QUADPACK}, p.110,
\begin{eqnarray}
   && f(x) = \frac{1}{|x\sp{2}+2x-2|\sp{1/2}} =
          \frac{1}{\sqrt{|(x-x\sb{p})(x-x\sb{m})|}},
\nonumber\\
    &&{\rm where}\quad\quad x\sb{p} = \sqrt{3}-1,\ x\sb{m} = -\sqrt{3}-1.
\label{eq:snglr3}
\end{eqnarray}
The values $\{f(x\sb{k});\ x\sb{k}\! =\! fl(x\sb{p})(1+k\varepsilon\sb{0});\
    \varepsilon\sb{0}\! =\! 2\sp{-53}| \ k\! =\! -20(1)20\}$
have been computed and compared with the exact ones.
Several spuriously equal output pair are noticed at pairs of assumed
neighbouring machine number arguments, which spoil the reliability of
the analysis.
\par
Increase of $\varepsilon\sb{0}$ to $\varepsilon\sb{0}=2\sp{-52}$ rules
out such spurious pairs (Fig.~1c).
\par
{\bf 5.2. Floating point comparisons.}
In the example~(\ref{eq:snglr3}), $fl(x\sb{p})$ may be either transferred
from a procedure to another one using the $fl\sb{52}(x\sb{p})$ value
which retains the most significant $52$ binary bits in the significand,
or it may be directly computed in CPU from the original expression
$x\sb{p} = \sqrt{3}-1$, in which case the processor stored
$fl\sb{64}(x\sb{p})$ value retains the most significant $64$ binary bits.
The IEEE 754 standard asks that
$fl\sb{64}(x\sb{p})${\tt .EQ.}$fl\sb{52}(x\sb{p})=${\tt .TRUE.}
None of the two compilers available to us (f77, C++ gcc) did obey to this
requirement.
\par
Since the use of floating point comparison was unavoidable, special
care was taken to exclude all the possible $fl\sb{52}$ to $fl\sb{64}$
comparisons.
\par
{\bf 5.3. Code optimization by the compiler.}
We might try to write a code in which use of $fl\sb{52}$ values is
always secured in comparison operations by asking the transfer of each
variable entering such operation to RAM/cache, followed by their transfer
back to CPU. In its search for "efficiency" increase, the code optimization
by the compiler finds such a trick "unnecessary", thus spoiling the output
correctness. The observation is not singular
\cite{kah2}.
\par
{\bf 5.4. The underflow threshold.}
On all but one computers at our disposal, the standard value
$u=2\sp{-1023}$ was found to be right. However, on one of the mentioned
SUN workstations the code crashed at $u=2\sp{-1023}$, while the value
$u=2\sp{-1022}$ was found to be OK.
\par
{\bf 5.5. Catastrophic precision loss in the neighbourhood of a non-zero
singular point.}
Two supplementary features present in figures 1b and 1c deserve consideration:
(i) The computed function values are different from the exact ones
\emph{starting with the most significant bit}.
(ii) The use of the primary QUADPACK expression listed in
Eq.~(\ref{eq:snglr3}) (data labelled "l53","l52") infringes the left-right
symmetry of the exact data.
\par
In spite of this severe precision loss due to cancellation by subtraction,
correct inferences based on the use of the unique features characterizing a
singular behaviour (paragraph 4.2) are still possible due to the fact that
the computed "wrong data" \emph{preserve\/} the ordering relationships
characteristic to the "true data".
\par
If a substitution of variable which moves the singularity to the origin is
possible, then all the difficulties enumerated at paragraphs 5.2, 5.3, and
5.5 are completely removed (Fig.~1d).
\section{Conclusions}
The boundary layer problem asks for accurate and reproducible diagnostics of
integrand $f(x)$ behaviour at the endpoints $a$ and $b$ of a finite
integration domain $[a, b]$.
\par
In the present paper we have discussed several critical issues which
dramatically influence code robustness and reliability.
Generalizations of the results previously reported in
\cite{AATN05}
allows significant improvement of code quality by:
\par
(i) Definition of smooth behaviour from an \emph{unrestricted\/} least
squares analysis over small (mesoscopic) neighbourhoods of the endpoints
$a$ and $b$. This secures the derivation of continuity criteria valid
\emph{everywhere\/} over the mesoscopic range where the analysis is done.
\par
(ii) Formulation of qualitative diagnostic criteria which reliably single
out the various kinds of integrand behaviour even under severely damaged
accuracy of the computed data.
\par
(iii) Derivation of accurate tests for reliable definition of the machine
epsilon with respect to the addition.
\par
(iv) Identification of the critical hardware and software features which could
spoil the correctness of the diagnostics by deviation from the IEEE 754
standard and code reformulation such as to become insensitive to such
drawback of the computing environment.
\par
The correct solution of the boundary layer problem is the first from a set
of hierarchically ordered problems the solutions of which should allow
\emph{a priori\/} Bayesian inferences on efficient and reliable mesh
generation within automatic adaptive quadrature.
The solution of such problems is planned to be discussed in subsequent
reports.
\par
{\small {\bf Acknowledgments.}
This research was partially financed within the CEEX Contract
CEx 05-D11-68/11.10.2005 in IFIN-HH and theme 09-6-1060-2005/2007 in LIT-JINR.
The authors acknowledge the financial support received
within the "Hulubei-Meshcheryakov" Programme, JINR order
No. 726/06.12.2005 for participation to MMCP 2006 Conference.}

\vfil\eject
\def\endpage{\hfill\eject}

\end{document}